\documentclass[12pt,a4paper]{article}
\usepackage{graphicx,psfrag,amsmath,amssymb,amsfonts,latexsym,color,bm}
\usepackage[english]{babel}
\usepackage[utf8]{inputenc}
\usepackage{slashed}
\usepackage{physics}
\usepackage{tikz}
\usepackage{pgfplots}
\pgfplotsset{compat=1.17}
\usepackage[margin=2.5cm]{geometry}
\usepackage{simplewick}
\usepgfplotslibrary{polar}
\begin{document}
\title{On the Sign of the Casimir Force Between Two Bodies Coupled to a
Dirac field} 
\author{Aitor Fern\'andez, C\'esar D.~Fosco and Guillermo Hansen\\
{\normalsize\it Centro At\'omico Bariloche and Instituto Balseiro}\\
{\normalsize\it Comisi\'on Nacional de Energ\'{\i}a At\'omica}\\
{\normalsize\it R8402AGP S.\ C.\ de Bariloche, Argentina.} }
\maketitle
\begin{abstract}
We apply the result of Kenneth and I.~Klich, Phys.\ Rev.\ Lett.\
\textbf{97}, 160401 (2006) to derive a theorem
for two objects coupled to a Dirac field. We demonstrate that a sufficient
condition for the Casimir interaction  to be {\em repulsive}, is
for their geometric configuration to be {\em odd} under the same
reflection operation considered in that work. 
\end{abstract}
\section{Introduction}\label{sec:intro}
The study of the relation between the properties of the Casimir 
effect~\cite{Casimir1948} and the geometry (i.e., shapes and positions) 
of the bodies involved is a captivating area of research, with significant 
implications to both experimental and theoretical physics. Understanding 
this relation is crucial for Casimir experiments; indeed, one could, for 
instance, be interested in discovering
the optimal geometric configuration to maximize the Casimir force, as this
could have applications in nanotechnology and microelectromechanical
systems (MEMS)~\cite{Chan2001}.  Additionally, identifying shapes that
exhibit stability under small deformations is vital for ensuring the
reliability and robustness of devices that exploit the Casimir 
effect~\cite{Rodriguez2011}.
Furthermore, investigating the impact of surface corrugations and other
irregularities on the Casimir force is essential for predicting its
behavior in real-world scenarios~\cite{Banishev2013}.

On the theoretical front, an important challenge lies in developing
effective methods to cope with the geometries that often arise in Casimir
effect studies. Perturbation theory, a commonly used tool in theoretical
physics, is generally not suitable for tackling non-trivial 
shapes~\cite{Emig2001}, except
in specific cases, like an expansion in powers of the deformation of the
geometry with respect of a highly symmetric one, which is exactly solvable.
A related but different approach is the derivative
expansion~\cite{Fosco2011}. 

This calls for the development of alternative theoretical frameworks and
computational techniques that can accurately model the Casimir interaction
for a wide range of geometric configurations.  In this context, there is an
important theoretical result regarding the relation between geometry and
the  Casimir force: a sufficient condition for this force between
two dielectric or conducting objects to be attractive is for the geometry of the system to have a reflection symmetry with respect to a
plane~\cite{Kenneth:2006}. In other words, when the function representing
the electromagnetic properties of the two media as a whole is invariant
(i.e., even) under the symmetry operation, the force between them is
attractive.

In this work, we apply that theorem to derive another one, this
time for objects coupled to a Dirac field.  The result may be phrased as
follows: a sufficient condition for the Casimir interaction to be {\em 
repulsive}, is for their geometric configuration to be {\em odd} under the 
same geometric operation considered in the bosonic case.

This paper is organized as follows: since our approach to the Dirac field
case relies heavily upon the scalar field result, we briefly review the 
latter in Sect.~\ref{sec:Scalar}, albeit in the notation and
conventions we afterwards use for the Dirac field. Besides, we focus here 
on the properties which are more relevant to the Dirac field case, which 
is dealt with in Sect.~\ref{sec:Dirac}, where we derive the result about 
the  sign of the force in that kind of system.  

In Sect.~\ref{sec:Conclusions} we present some remarks about the general result
and its application, and a brief summary of our conclusions.

\section{Scalar field}\label{sec:Scalar}
We shall begin by introducing the main concepts and definitions for the
case of a real scalar field, briefly reviewing the known result for the
nature of the force, in terms of our approach, notation, and conventions.
We deal with a vacuum massless real scalar field $\varphi(x)$, in $d+1$
space-time dimensions, in the presence of two objects, denoted by $L$ and
$R$.
To obtain the Casimir interaction energy, we shall work in the context of
the functional integral framework, using the imaginary-time formalism;
hence, it is convenient to introduce the system in terms of its Euclidean
action.  This object is a functional ${\mathcal S} = {\mathcal S}(\varphi;
U)$ where $U$ denotes a function that accounts for the relevant properties
of the two bodies.  We assume ${\mathcal S}$ to have the structure:
\begin{equation}
	{\mathcal S} =  {\mathcal S}^{(0)} \,+\, {\mathcal S}^{(I)} \;,
\end{equation}
where ${\mathcal S}^{(0)}$ is the free action,
\begin{equation}
	{\mathcal S}^{(0)}(\varphi) \,=\,\frac{1}{2} \,\int_x  
\partial_\mu\varphi(x)  \partial_\mu \varphi(x) \;,  
\end{equation}  
and we have adopted a shorthand notation for the integration, namely,
$\int_x \ldots \equiv \int d^{d+1}x \ldots $. The Einstein convention on
summation over repeated indices has been assumed. Indices from the
middle of the Greek alphabet run from $0$ to $d$, and they correspond to
space-time coordinates. We shall use $x=(x_0,x_1,\ldots,x_d)$ to denote
these  coordinates, $x_0$ is the imaginary time (the speed of light,
$c \equiv 1$), and ${\mathbf x}=(x_1,\ldots,x_d)$ the spatial coordinates.
The ${\mathcal S}^{(I)}$ term, on the other hand, deals with the coupling
between the field and the two bodies, and is given by:
\begin{equation}
{\mathcal S}^{(I)}(\varphi ; U) \,=\, 
\int_{x_0,x'_0, {\mathbf x}}  \varphi(x_0,{\mathbf x})
U(x_0-x'_0;{\mathbf x}) \varphi(x'_0,{\mathbf x}) \;.
\end{equation}
Working with the Fourier transformed versions, with respect to the time 
coordinate, of the objects involved~\footnote{We use the notation: 
$\int_{\not\omega}\ldots \equiv \int 
\frac{d\omega}{2\pi} \ldots$}:
\begin{equation}
\varphi(x_0,{\mathbf x}) =\int_{\not\omega} e^{i \omega x_0} 
\widetilde{\varphi}(\omega, {\mathbf x})\;,\;\;
U(x_0;{\mathbf x}) =\int_{\not\omega} e^{i \omega x_0} 
\widetilde{U}(\omega , {\mathbf x}) \;,
\end{equation}
 the action decomposes into an integral over 
independent modes, one for each frequency. Namely,
\begin{equation}
{\mathcal S}(\varphi; U) = \int_0^\infty \frac{d\omega}{2 \pi} \, 
{\mathcal S}_\omega(\widetilde{\varphi}^*, \widetilde{\varphi};
\widetilde{U})
\end{equation}
where ${\mathcal S}_\omega \,=\, {\mathcal S}_\omega^{(0)} \,+\, {\mathcal
S}_\omega^{(I)}$, such that:
\begin{align}
{\mathcal S}_\omega^{(0)} \,=\, \int_{\mathbf x}\, 
\widetilde{\varphi}^*(\omega ,{\mathbf x}) \big(-{\mathbf \nabla}^2 + \omega^2 \big)
\widetilde{\varphi}(\omega ,{\mathbf x}) \;,\;\;\;
{\mathcal S}_\omega^{(I)}\,=\, \int_{\mathbf x} \, 
\widetilde{\varphi}^*(\omega ,{\mathbf x}) \,\widetilde{U}(\omega, {\mathbf x})
\, \widetilde{\varphi}(\omega ,{\mathbf x}) \;,
\end{align}
with $^*$ denoting complex conjugation.
Note that, $\widetilde{U}(\omega, {\mathbf x}) = \omega^2 \chi(\omega, 
{\mathbf x})$, where $\chi$ is the susceptibility.

In order to obtain the vacuum energy, we introduce the Euclidean vacuum 
transition amplitude ${\mathcal Z}$, and the effective action $\Gamma$:
\begin{align}
{\mathcal Z}(U) &=\; e^{- \Gamma(U)} \,=\, 
\int {\mathcal D}\varphi \, e^{- {\mathcal S}(\varphi; U)} \,=\, 
\prod_{\omega =0}^{\infty} {\mathcal Z}_\omega(\widetilde{U}) \;, \nonumber\\
{\mathcal Z}_\omega(\widetilde{U}) &\equiv\, \int  {\mathcal 
D}\widetilde{\varphi}^* {\mathcal D}\widetilde{\varphi}\;
e^{ - {\mathcal S}_\omega } \;.
\end{align}
The vacuum energy $E(U)$, measured with respect to the empty space 
vacuum, is then given by:
\begin{equation}
	E(U) \;=\; \lim_{T \to \infty} \frac{1}{T} \big[ \Gamma_T(U)  - 
	\Gamma_T(0)\big]
\end{equation}
where $\Gamma_T$ is formally identical to $\Gamma$, but defined for a 
finite time interval. We then see that 
\begin{equation}
E(U) \;=\; -  \int_0^\infty \frac{d\omega}{2 \pi} 
\; \log \Big[\frac{{\mathcal Z}_\omega(\widetilde{U})}{{\mathcal 
Z}_\omega(0)} \Big] \;.
\end{equation}

To proceed, we now deal with the properties of the two objects, $L$ and 
$R$. 
We shall assume that:
\begin{equation}
\widetilde{U}(\omega,{\mathbf x}) \,=\, \widetilde{U}_L(\omega,{\mathbf x}) \,+\,
\widetilde{U}_R(\omega , {\mathbf x}) \,,
\end{equation}
such that, denoting by $L$ and $R$ the support of $U_L$ and $U_R$, respectively,
one has  $L\, \cap \, R \,=\, \emptyset$.  
Besides, to be consistent with the assumption of the susceptibility being
proportional to $\widetilde{U}$, both $\widetilde{U}_L$ and $\widetilde{U}_R$ must 
be real, non-negative functions. 

Taking into account those assumptions, we may produce an equivalent 
representation of the functional integral for ${\mathcal Z}_\omega$, this time 
introducing complex auxiliary fields $\lambda_L$ and $\lambda_R,$ with 
domains $L$ and $R$, respectively, and their Fourier transforms. This
amounts to replacing ${\mathcal S}_\omega^{(I)}$ by 
${\mathcal S}_\omega^{({\rm aux})}$, so that
\begin{equation}
{\mathcal Z}_\omega(\widetilde{U}) \;=\; 
\int {\mathcal D}\widetilde{\varphi}^* {\mathcal D}\widetilde{\varphi}\,
{\mathcal D}\widetilde{\lambda}_L^* {\mathcal D}\widetilde{\lambda}_L \,
{\mathcal D}\widetilde{\lambda}_R^* {\mathcal D}\widetilde{\lambda}_R \;
	e^{- \big( {\mathcal S}_\omega^{(0)} + {\mathcal S}_\omega^{({\rm
	aux})} \big)} \;,
\end{equation}
with
\begin{align}
{\mathcal S}_\omega^{({\rm aux})} &=\, 
\int_{{\mathbf x} \in L }\widetilde{\lambda}_L^*({\mathbf x})
\widetilde{U}_L({\mathbf x}) \widetilde{\lambda}_L({\mathbf x}) \,+\,
\int_{{\mathbf x} \in R }\widetilde{\lambda}_R^*({\mathbf x}) \widetilde{U}_R({\mathbf x}) 
\widetilde{\lambda}_R({\mathbf x})  \nonumber\\
& - i \int_{{\mathbf x} \in L } \big[ 
\widetilde{\lambda}_L^*({\mathbf x}) \widetilde{U}_L({\mathbf x}) 
\widetilde{\varphi}({\mathbf x}) 
+ \widetilde{\varphi}^*({\mathbf x}) \widetilde{U}_L({\mathbf x})
\widetilde{\lambda}_L({\mathbf x}) \big] 
\nonumber\\emig
& - i \int_{{\mathbf x} \in R } \big[ 
\widetilde{\lambda}_R^*({\mathbf x}) \widetilde{U}_R({\mathbf x}) 
\widetilde{\varphi}({\mathbf x}) 
+ \widetilde{\varphi}^*({\mathbf x}) \widetilde{U}_R({\mathbf x})
\widetilde{\lambda}_R({\mathbf x}) \big] \,.
\end{align}
For the sake of clarity, we have omitted writing explicitly the 
dependence on $\omega$ of the objects above.

To proceed, we integrate out $\widetilde{\varphi}^*, \widetilde{\varphi}$.
The result of this step may be put in the following form:
\begin{equation}
	\frac{{\mathcal Z}_\omega(\widetilde{U})}{{\mathcal
	Z}_\omega(0)}\;=\; \int  
{\mathcal D}\widetilde{\lambda}_L^* {\mathcal D}\widetilde{\lambda}_L \,
{\mathcal D}\widetilde{\lambda}_R^* {\mathcal D}\widetilde{\lambda}_R \;
e^{- {\mathcal S}_{\rm eff}} 
\end{equation}
\begin{equation}
{\mathcal S}_{\rm eff} \;=\; \sum_{A,B=L}^R 
\int_{{\mathbf x} \in A, {\mathbf x'} \in B} 
\widetilde{\lambda}_A^*({\mathbf x}) 
	K_{A B}({\mathbf x},{\mathbf x'})
\widetilde{\lambda}_{B}({\mathbf x'}) \;,
\end{equation}
where we have introduced the coordinate-space kernels:
\begin{equation}
K_{AB}({\mathbf x},{\mathbf x'}) \;=\; \delta_{AB} \, \delta({\mathbf
x}-{\mathbf x'}) \widetilde{U}_A({\mathbf x}) \,+\, 
\widetilde{U}_A({\mathbf x}) G^{(0)}_{AB}({\mathbf x}-{\mathbf x'})
\widetilde{U}_B({\mathbf x'}) \;,
\end{equation}
with 
\begin{equation}
G^{(0)}_{AB}({\mathbf x}-{\mathbf x'}) \,\equiv\,
\langle{\mathbf x}|\hat{G}^{(0)}_{AB}|{\mathbf x'}\rangle \,=\, 
\langle {\mathbf x}|\big(-\nabla^2 + \omega^2  \big)^{-1}
|{\mathbf x'}\rangle \;,\;\;{\mathbf x} \in A, \; {\mathbf x'}\in B \;. 
\end{equation}

We now introduce some notation to distinguish an operator from its
coordinate kernel. For instance, to the operator $\widehat{K}_{AB}$ 
corresponds the kernel $K_{AB}({\mathbf x},{\mathbf x'}) = \langle 
{\mathbf x}|\widehat{K}_{AB} |{\mathbf x'}\rangle$.  This is an operator 
acting from the space of functions $H_A = L^2(A)$ the space of square 
integrable functions with support $A$ into those with support $B$. 
We will adopt the convention that indices like $A$, $B$, are assumed to 
run over the values that label the two objects, namely, $L$ or $R$.

Using standard formulae for the evaluation of Gaussian integrals, and 
noting that both $\widehat{K}_{LL}$ and $\widehat{K}_{RR}$ are 
non-singular, we get:
\begin{equation}
\frac{{\mathcal Z}_\omega(\widetilde{U})}{{\mathcal Z}_\omega(0)}\;=\; 
[{\rm det}(\widehat{K}_{LL})]^{-1} \, [{\rm det}(\widehat{K}_{RR})]^{-1} \,
\Big( {\rm det}\big[ \widehat{I}_L - (\widehat{K}_{LL})^{-1} \widehat{K}_{LR}
(\widehat{K}_{RR})^{-1} \widehat{K}_{RL} \big] \Big)^{-1} \;,
\end{equation}
where $\widehat{I}_A$ is the identity operator for functions with support
on $A$.

Finally, using the assumption that each function $\widetilde{U}_A$ is nonnegative, its operator $\widehat{\widetilde{U}}_A$, such that 
$\langle {\mathbf x}|\widehat{\widetilde{U}}_A|{\mathbf x'}\rangle =
\widetilde{U}_A({\mathbf x}) \delta({\mathbf x}-{\mathbf x'})$ has an 
inverse.

Then, introducing
\begin{equation}
\widehat{T}_A = \big( \widehat{I}_A + \widehat{\widetilde{U}}_A
\hat{G}^{(0)}_{AA})^{-1} \widehat{\widetilde{U}}_A \;,
\end{equation}
we see that:
\begin{equation}
\frac{{\mathcal Z}_\omega(\widetilde{U})}{{\mathcal Z}_\omega(0)} \;=\;
({\rm det}\widehat{K}_{LL})^{-1} \, ({\rm det}\widehat{K}_{RR})^{-1} \; 
\big( {\rm det}[ \widehat{I}_L - \widehat{T}_L 
	\hat{G}^{(0)}_{LR} \widehat{T}_R \hat{G}^{(0)}_{RL}] 
\big)^{-1} \;.
\end{equation}

Thus, subtracting terms that survive when the two bodies are infinitely far
apart, and therefore do not contribute to the interaction energy, we get:
\begin{equation}
E_I \;=\; \int_0^\infty \frac{d\omega}{2\pi} \, {\rm log} {\rm det}[
\widehat{I}_L - \widehat{T}_L \hat{G}^{(0)}_{LR} \widehat{T}_R \hat{G}^{(0)}_{RL}] \;=\; 
\int_0^\infty \frac{d\omega}{2\pi} 
\, {\rm Tr} {\rm log}[ \widehat{I}_L - \widehat{T}_L
\hat{G}^{(0)}_{LR} \widehat{T}_R \hat{G}^{(0)}_{RL}] \;.
\end{equation}
This is, in our conventions, equivalent to Eq.(5) of
Ref.~\cite{Kenneth:2006}, which is used to prove the theorem on the 
attractive nature of the force when there is a reflection symmetry. 
Up to now, no assumption about the existence of that symmetry in the system
has been made. Let us now assume the existence of such a symmetry.
Assuming $x_d$ is the coordinate that changes under the reflection,
and denoting the transformation by a function ${\mathcal P}$, we have, at the
level of the coordinates,
${\mathbf x} \to {\mathcal P}({\mathbf x})$, where, for ${\mathbf
x}=(x_1,x_2, \ldots, x_d)$, ${\mathcal P}({\mathbf x}) = (x_1,x_2, \ldots,
x_{d-1},a- x_d)$. With this convention, the reflection plane is
$x_d=\frac{a}{2}$, the region invariant under the transformation.
On has ${\mathcal P}^2 = I$, and that this operation is linearly
represented on the space of fields by a unitary and self-adjoint operator
$\widehat{\mathcal P}$:
\begin{equation}
\widehat{\mathcal P} |{\mathbf x}\rangle \,=\, |{\mathcal
	P}({\mathbf x})\rangle \;.
\end{equation}
The action of $\widehat{\mathcal P}$ on functions $\psi({\mathbf x})$ 
of $H= H_L \oplus H_R$ is thus 
\begin{equation}
\langle {\mathbf x}| \widehat{\mathcal P} |\psi\rangle \,=\,
	\langle {\mathcal P}({\mathbf x})|\psi\rangle \,=\, \psi({\mathcal
	P}({\mathbf x})) \;.
\end{equation}

Besides, ${\mathcal P}$ interchanges $L$ and $R$:  ${\mathcal P}(L) = R$,
and  ${\mathcal P}(R) = L$,  (${\mathcal P}(A)$ denotes the  image of $A$).
This property is inherited in the functional spaces: $\widehat{\mathcal
P}(H_L) = H_R$, and $\widehat{\mathcal P}(H_R) = H_L$. 

Reflection symmetry implies that~\footnote{We recall that
$\widehat{\mathcal P}^\dagger = \widehat{\mathcal P}$.}: 
\begin{equation}
\widehat{\widetilde{U}}_R \,=\, \widehat{\mathcal P}
\widehat{\widetilde{U}}_L \widehat{\mathcal P}^\dagger \;,\;\;
\widehat{I}_R \,=\, \widehat{\mathcal P} \widehat{I}_L \widehat{\mathcal
	P}^\dagger \;,\;\;
\widehat{\tilde{G}}^{(0)}_{RR} \,=\, \widehat{\mathcal P}\widehat{\tilde{G}}^{(0)}_{LL}
 \widehat{\mathcal P}^\dagger \;, \;\; \widehat{\tilde{G}}^{(0)}_{LR} \,=\, 
\widehat{\mathcal P}\widehat{\tilde{G}}^{(0)}_{RL} \widehat{\mathcal P}^\dagger \;,
\end{equation}
and as a consequence: $\widehat{\mathcal P} \widehat{\widetilde{U}} 
\widehat{\mathcal P}^\dagger = \widehat{\widetilde{U}}$. 

These properties allow one to prove that
\begin{equation}\label{eq:eisq}
E_I \;=\; \int_0^\infty \frac{d\omega}{2\pi} \, {\rm log} \big[{\rm 
det}\big(
	\widehat{I}_L - \widehat{Y}^2 \big) \big] \;,\;\;
\widehat{Y}	\,\equiv\, \sqrt{\widehat{T}_L} \hat{G}^{(0)}_{LR} 
\widehat{\mathcal P} \sqrt{\widehat{T}_L}  \;.
\end{equation}
Following~\cite{Kenneth:2006} one can use (\ref{eq:eisq}) to find the sign 
of $F$, the force along the $x_d$ direction. First, $E_I$ is expressed in 
terms of the eigenvalues $\lambda_n$ of the Hermitian operator 
$\widehat{Y}$:
\begin{align}
	E_I&=\; \int_0^\infty \frac{d\omega}{2\pi} \,\sum_n  {\rm log} ( 1 - 
	\lambda_n^2) \nonumber\\
	F & =\; - \partial_a E_I \,=\, 2 \int_0^\infty \frac{d\omega}{2\pi} 
	\,\sum_n  \frac{\lambda_n \partial_a\lambda_n}{1 - 
	\lambda_n^2} \;.
\end{align}
The sign of the force is then that of the product $\lambda_n 
\partial_a\lambda_n$ (since $\lambda_n^2 < 1$ $\forall n$).

Introducing $|\phi_n\rangle$, the normalized eigenvectors of 
$\widehat{Y}$, $\lambda_n\,=\, \langle \phi_n | \widehat{Y} 
|\phi_n \rangle$. Thus,
\begin{align}
	\lambda_n \partial_a\lambda_n \,=\, \langle \phi_n | \widehat{Y} 
	|\phi_n \rangle  \, \langle \phi_n | \partial_a \widehat{Y} 
	|\phi_n \rangle 
\end{align}
where the Hellmann-Feynman theorem~\cite{Hellmann1937,Feynman1939} has been 
applied. 

Since $\widehat{T}$ is independent of $a$, it is convenient to  
introduce $|\psi_n \rangle \equiv \sqrt{\widehat{T}_L}|\phi_n\rangle$, 
and \mbox{$\widehat{Z} \,=\,  \hat{G}^{(0)}_{LR} \widehat{\mathcal P}$} to 
obtain:
\begin{align}
	{\rm sign}(\lambda_n \partial_a\lambda_n) \,=\, {\rm sign} \big[ 
	\langle \psi_n | \widehat{Z} 
	|\psi_n \rangle \big] \, {\rm sign} \big[\langle \psi_n | \partial_a 
	\widehat{Z} 
	|\psi_n \rangle \big] \;.
\end{align}

In~\cite{Kenneth:2006} it is shown that $\widehat{Z}$ is 
positive, while $\partial_a\widehat{Z}$ is  negative. This, of 
course has the consequence that ${\rm sign}(\lambda_n \partial_a\lambda_n) 
= -1$, for all 
$n$, and the force is then negative, i.e., attractive. There is an 
important remark we wish to make here: in order to have a negative force, 
it is sufficient to have opposite signs for $\langle \psi_n | 
\widehat{Z} |\psi_n \rangle$ and $ \langle \psi_n | \partial_a 
\widehat{Z} |\psi_n \rangle$, for each $n$. To be more precise, the signs 
have to be opposite for the same $n$, but the sign of each factor could 
depend on $n$, as long as the sign of the product is $(-1)$ for all $n$. 
As we shall see, this remark is relevant to the Dirac field case:

We conclude this Section by recalling the assumption for 
the functions $\widetilde{U}_A$ to be positive definite. 
Note also that both the symmetry and positive definiteness can be 
achieved, with no change in the theorem, by an $\widetilde{U}$ which has a 
matrix structure, assuming a multicomponent scalar field. By this, we mean 
that
\begin{equation}
	{\mathcal S}^{(0)}(\varphi) \,=\,\frac{1}{2} \,\int_x  
	\partial_\mu\varphi^{(\alpha)}(x)  
	\partial_\mu \varphi^{(\alpha)}(x) \;,  
\end{equation}
with $\alpha = 1, \ldots, N$, and 
\begin{equation}
	{\mathcal S}^{(I)}(\varphi ; U) \,=\, 
	\int_{x_0,x'_0, {\mathbf x}}  \varphi^{(\alpha)}(x_0,{\mathbf x})
	U^{(\alpha \beta)}(x_0-x'_0;{\mathbf x}) \varphi^{(\beta)}(x'_0,{\mathbf 
	x}) \;,
\end{equation}
where $U$ is  positive.

As in the single-component case, we use the same time Fourier 
representation, and introduce a column vector $\widetilde{\phi} \equiv 
(\widetilde{\varphi}^{(\alpha)})$, as well as  a 
matrix $\widetilde{\mathbb U}= [\widetilde{U}^{(\alpha \beta)}]$. We again 
have a decoupling of the action into independent terms, one for each 
frequency:
\begin{align}
	{\mathcal S}_\omega^{(0)} \,=\, \int_{\mathbf x}\, 
	\widetilde{\phi}^\dagger(\omega ,{\mathbf x}) \big(-{\mathbf 
	\nabla}^2 + 
	\omega^2 \big)\widetilde{\phi}^\dagger(\omega ,{\mathbf x}) \;,\;\;\;
	{\mathcal S}_\omega^{(I)}\,=\, \int_{\mathbf x} \, 
	\widetilde{\phi}^\dagger(\omega ,{\mathbf x}) \,\widetilde{\mathbb 
	U}(\omega, 
	{\mathbf x})
	\, \widetilde{\phi}(\omega ,{\mathbf x}) \;.
\end{align}

In this case, a generalization of the single-component one, essentially 
the same proof allows one to conclude that the force is attractive under 
the assumption of the parity:
$\widehat{\widetilde{\mathbb U}}_R \,=\, \widehat{\mathcal P}
\widehat{\widetilde{\mathbb U}}_L \widehat{\mathcal P}^\dagger$. 
The $\widehat{T}_A$ operators now also have a matrix structure, but the 
rest of the conclusion that when there is symmetry the force is attractive, 
remains unaltered.  

\section{Dirac field}\label{sec:Dirac}
Let us now deal with a vacuum massless Dirac field $\psi$, $\bar{\psi}$, in
$d+1$ space-time dimensions, in the presence of two objects, $L$ and $R$, which
impose boundary conditions on the field fluctuations. The regions are
assumed to be disjoint, as in the scalar case. 

We shall first relate the calculation of the vacuum energy to the case of a
real scalar field, and then study the role of the reflection transformation
in this case.

The Euclidean action is now a functional ${\mathcal S} = {\mathcal
S}(\bar{\psi},\psi;V)$ where $V$ denotes a function that accounts for the
two bodies. We also assume ${\mathcal S}$ to have the structure:
\begin{equation}
	{\mathcal S} =  {\mathcal S}_0 \,+\, {\mathcal S}_I \;,
\end{equation}
where ${\mathcal S}_0$ denotes the free action for the Dirac field,
\begin{equation}
{\mathcal S}_0(\bar{\psi},\psi) \;=\; 
\int_x  \bar{\psi}(x)\not\!\partial \, \psi(x) \;. 
\end{equation}  

Dirac's $\gamma$ matrices are assumed to be Hermitian, 
$\gamma_\mu^\dagger = \gamma_\mu$, and to satisfy the Clifford algebra 
$\{ \gamma_\mu \,,\, \gamma_\nu \} = 2 \delta_{\mu\nu}$.  
Also, $\not\!\partial \equiv \gamma_\mu \partial_\mu$.

The ${\mathcal S}_I$ term, on the other hand, is given by:
\begin{equation}
{\mathcal S}_I(\bar{\psi}, \psi;V) \,=\, 
\int_{x_0,x'_0, {\mathbf x}}  \bar{\psi}(x_0,{\mathbf x})
V(x_0-x'_0;{\mathbf x}) \psi(x'_0,{\mathbf x}) \;.
\end{equation}

Introducing time Fourier transformations,
\begin{equation}
\psi(x_0,{\mathbf x}) =\int_{\not\omega} e^{i \omega x_0} 
\tilde{\psi}(\omega, {\mathbf x})\;,\;\;\bar{\psi}(x_0,{\mathbf x}) = 
\int_{\not\omega} 
e^{- i \omega x_0} \bar{\tilde{\psi}}(\omega,{\mathbf x})\;,\;\;
V(x_0;{\mathbf x}) =\int_{\not\omega} 
e^{i \omega x_0} \widetilde{V}(\omega , {\mathbf x})\;,
\end{equation}
the action decomposes again 
into an integral over independent modes
\begin{equation}\label{eq:dirop}
{\mathcal S} \,=\, \int_{\not \omega} \, {\mathcal S}_\omega \;,\;\;\:
{\mathcal S}_\omega \,=\, \int_{\mathbf x} \, 
\bar{\tilde{\psi}}(\omega ,{\mathbf x}) \big[ \not \! \nabla + i \gamma_0 
\omega + 
\widetilde{V}(\omega, {\mathbf x}) \big] \, \tilde{\psi}(\omega ,{\mathbf 
x})	\;.
\end{equation}

The properties of the two objects,  are 
again determined by the form of $V$:
\begin{equation}
\widetilde{V}({\mathbf x}) \,=\, \widetilde{V}_L({\mathbf x}) \,+\, 
\widetilde{V}_R({\mathbf x}) \,.
\end{equation}

A crucial difference between this case and the bosonic one is that the 
$\widetilde{V}_A$ functions do not need to be positive. Indeed, on physical 
ground, the bosonic $\widetilde{U}$ plays the role of the square of a mass, 
unlike 
$\widetilde{V}$.
We do assume, however, that the $\widetilde{V}_A$ functions do not vanish 
inside the respective regions, $A = L, R$, irrespective of their sign.

Then we see that, as in the bosonic case:
\begin{equation}
	e^{-\Gamma(V)} \;=\; \frac{{\mathcal Z}(V)}{{\mathcal Z}(0)} \;, 
\end{equation}
where 
\begin{equation}
	{\mathcal Z}(V) \,=\, \int {\mathcal D}\bar{\psi} {\mathcal D}\psi
	\, e^{-{\mathcal S}(\bar{\psi}, \psi;V)} \;.
\end{equation}
Thus,
\begin{equation}\label{eq:evf}
E(V) \,=\, - \big[\frac{\Gamma(V)}{T}]_{T\to\infty} \,=\, 
- \, \int_{\not \omega} \log \frac{\det\big(\not \! \nabla + i \gamma_0 
\omega 
+ \widetilde{V}\big)}{\det \big(\not \! \nabla + i \gamma_0 
\omega \big)} \;.
\end{equation}
Here, we have used the notation $\not \!\! \nabla \equiv \gamma_j 
\partial_j$, and the determinant is meant to be over functional as well as 
Dirac space.
In (\ref{eq:evf}), we take advantage of the fact that the energy is real, 
and that the determinant of the adjoint of an operator is the conjugate of 
the determinant of the original operator, to make
\begin{align}\label{eq:evf1}
	E(V) \,=\, E^*(V) \,=\,
	- \, \int_{\not \omega} \log \frac{\det\big[ -\not \! \nabla - i 
	\gamma_0 \omega 	+ \widetilde{V}\big]}{\det \big[ -\not \! 
	\nabla - i \gamma_0 \omega 
		\big]} \;,
\end{align}
and then
\begin{align}\label{eq:evf2}
	E(V) &=\, - \frac{1}{2} 
	\int_{\not \omega} \log \frac{\det\big[ (-\not \! \nabla - i \gamma_0 
	\omega 	+ \widetilde{V}) (\not \! \nabla + i \gamma_0 \omega 	+ 
	\widetilde{V})\big]}{\det\big[ (-\not \! \nabla - i \gamma_0 \omega) 
	(\not \! 	\nabla + i \gamma_0 \omega)\big]} \nonumber\\
	 &=\, - \frac{1}{2} \int_{\not \omega} \log \frac{\det\big[- \nabla^2  
	 + \omega^2 + \widetilde{V}^2- (\not \! \nabla 
	 \widetilde{V})\big]}{\det (-\nabla^2 + \omega^2)}\;.
\end{align}
This expression is identical {\em except for an important global factor 
$(-\frac{1}{2})$} to the one of the multi-component scalar field considered 
in the previous Section. 
The number of components depends, of course, on the dimension of the 
representation of the Clifford algebra generated by the Dirac matrices.

In the mapping to the scalar field case, we make the 
identification:
\begin{equation}\label{eq:ident}
	\widetilde{\mathbb{U}} \;=\; \widetilde{V}^2- (\not \! \nabla 
	\widetilde{V}) \;.
\end{equation}
Hence, the resulting $\widetilde{\mathbb U}$ 
is Hermitian and has support on the same regions $L$ and $R$ defined by 
$\widetilde{V}$.

Regarding the positive definiteness of $\widetilde{\mathbb U}$ it is 
determined by the relative magnitude of the norms of
$\widetilde{V}^2$ and $(\not\!\!\nabla \widetilde{V})$. We shall assume 
$\widetilde{V}$, or what is the same, each one of its components 
$\widetilde{V}_L$ and $\widetilde{V}_R$ are such that the $	
\widetilde{\mathbb{U}}$ is positive. 

In the Dirac case, we then have:
\begin{equation}\label{eq:eid}
	E_I \;=\; -\frac{1}{2} \int_{\not\omega} \, \log \det \, [
	\widehat{I}_L - \widehat{T}_L \hat{G}^{(0)}_{LR} \widehat{T}_R 
	\hat{G}^{(0)}_{RL}] \;=\; -\frac{1}{2} \int_{\not\omega} \,
	\, {\rm Tr}  \log \, [ \widehat{I}_L - \widehat{T}_L
	\hat{G}^{(0)}_{LR} \widehat{T}_R \hat{G}^{(0)}_{RL}] \;,
\end{equation}
with $\hat{G}^{(0)}_{AB}$ diagonal in Dirac space, and $\widehat{T}_A$ 
identical to 
the ones of the (multicomponent) bosonic case, with	
$\widetilde{\mathbb{U}}_A$ determined by $\widetilde{V}_A$, as in 
(\ref{eq:ident}). 
Note the important difference posed by the global $(-\frac{1}{2})$ factor, 
which is crucial for determining the sign of the force.

Let us now consider the issue of symmetry under reflection. Albeit the 
conclusion will be identical we need to distinguish between two cases: 
$d=1$ and $d > 1$:
\subsection{$d = 1$}
For $d = 1$ it is quite clear that if
\begin{equation}\label{eq:anti}
\widehat{\mathcal P} \widetilde{V}_L \widehat{\mathcal P}^\dagger \, = \, 	
-  \widetilde{V}_R \;\;,\;\;\; 
\widehat{\mathcal P} \widetilde{V}_R \widehat{\mathcal P}^\dagger \, = \, 	
-  \widetilde{V}_L \;\; \Rightarrow \;
\widehat{\mathcal P} \widetilde{V} \widehat{\mathcal P}^\dagger \, = \, 	
-  \widetilde{V} \;,
\end{equation}
then:
$\widehat{\widetilde{\mathbb U}}_R \,=\, \widehat{\mathcal P}
\widehat{\widetilde{\mathbb U}}_L \widehat{\mathcal P}^\dagger$. 
Thus, under this antisymmetry assumption for $\widetilde{V}$, we may apply 
the same arguments of the bosonic case for an $U$ which is parity 
symmetric. Therefore:
${\rm sign}(\lambda_n \partial_a\lambda_n) \,=\, -1$, $\forall n$.

Thus, because of the $-\frac{1}{2}$ in (\ref{eq:eid}):
\begin{equation}
	{\rm sign}(F) \,=\, + 1 \;,
\end{equation}
which is a repulsive Casimir interaction. Again, note  that the conclusion 
does not change if the Dirac field has many flavors and therefore the 
previous antisymmetry is realized for the resulting Hermitian matrices 
$\widetilde{\mathbb V}_L$ and  $\widetilde{\mathbb V}_R$.

\subsection{$d > 1$}
For $d > 1$ we postulate the same anti-symmetry as for $d=1$, namely, that 
(\ref{eq:anti}) holds. This yields, for $\widehat{\widetilde{\mathbb U}}$:
\begin{equation}
\widehat{\mathcal P} \gamma_d \widetilde{\mathbb U}_L \gamma_d 
\widehat{\mathcal P}^\dagger 
\, = \, \widetilde{\mathbb U}_R \;\;,\;\;\; 
\widehat{\mathcal P} \gamma_d \widetilde{\mathbb U}_R \gamma_d 
\widehat{\mathcal P}^\dagger \, = \, \widetilde{\mathbb U}_L \;.
\end{equation}
Since $\widehat{\mathcal P} \gamma_d \widetilde{T}_L 
\gamma_d \widehat{\mathcal P}^\dagger 
\, = \, \widetilde{T}_R$ we find:
\begin{equation}\label{eq:eisda}
	E_I \;=\; - \frac{1}{2} \int_{\not \omega}
	\, {\rm log} \big[{\rm det}\big(
	\widehat{I}_L - \widehat{Y_d}^2 \big) \big] \;,\;\;
	\widehat{Y_d}	\,\equiv\, \sqrt{\widehat{T}_L} \hat{G}^{(0)}_{LR} 
\widehat{\mathcal P} \gamma_d  \sqrt{\widehat{T}_L}  \;.
\end{equation}

Now, following the same procedure as for the real scalar field 
case, we introduce $|\phi_n\rangle$, now the normalized eigenvectors of 
$\widehat{Y}_d$ (note that they cannot be chosen to be, simultaneously, 
eigenvectors of $\gamma_d$)  and arrive to the study of the sign of
\begin{align}
\langle \phi_n | \widehat{Z}_d |\phi_n \rangle  \, \langle \phi_n | 
\partial_a \widehat{Z}_d|\phi_n \rangle \;, 
\end{align}
with \mbox{$\widehat{Z}_d \,=\, \widehat{Z}\,\gamma_d$}, where 
$\widehat{Z} = \hat{G}^{(0)}_{LR}\widehat{\mathcal P}$.

Again, we introduce $|\psi_n \rangle \equiv 
\sqrt{\widehat{T}_L}|\phi_n\rangle$, 
and also assume that a representation has been used for the Dirac matrices 
such that $\gamma_d$ is diagonal, each block having $+1$ or $-1$ as 
eigenvalue. Denoting by $|\psi_n^{(\pm)} \rangle$ the component of the 
respective block of 
$\gamma_d$, we get:
\begin{align}
{\rm sign}(\lambda_n \partial_a\lambda_n) & =\, {\rm sign} \big[ 
\langle \psi_n^{(+)} | \widehat{Z} |\psi_n^{(+)} \rangle - 
\langle \psi_n^{(-)} | \widehat{Z} |\psi_n^{(-)} \rangle\big] \nonumber\\
& \times \, 
 {\rm sign} \big[ 
\langle \psi_n^{(+)} | \partial_a\widehat{Z} |\psi_n^{(+)} \rangle - 
\langle \psi_n^{(-)} | \partial_a\widehat{Z} |\psi_n^{(-)} \rangle\big]
 \;.
\end{align}
Then we notice that the structure of each one of the averages above is as 
follows (see~\cite{Kenneth:2006}):
\begin{equation}
\langle \psi_n^{(\pm)} | \widehat{Z} |\psi_n^{(\pm)}\rangle \,\equiv\,
I_n^{(\pm)}(a) \,=\, \int \frac{d^{d-1}\mathbf{k}_\perp}{(2\pi)^{d-1}} \, 
\frac{e^{-a\sqrt{\mathbf{k}_\perp^2 + 
	\omega^2}}}{2\sqrt{\mathbf{k}_\perp^2 + \omega^2}} 
	f_n^{(\pm)}(\mathbf{k}_\perp,\omega) \,,
\end{equation}
where $f_n^{(\pm)}$ is a positive function (determined by 
$|\phi_n\rangle$) which form we do not need here.

Then,
\begin{equation}
{\rm sign}(\lambda_n \partial_a\lambda_n) \,=\, 
{\rm sign} \big[I_n^{(+)}(a) - I_n^{(-)}(a)\big] 
\times {\rm sign} \big[ \partial_a I_n^{(+)} - \partial_a I_n^{(-)}\big] 
\;.
\end{equation}
Assuming  $I_n^{(+)}(a) > I_n^{(-)}(a)$, and noting that, if that is 
the case, then, for $\varepsilon > 0$,
\begin{equation}
	I_n^{(+)}(a + \varepsilon) \,-\, I_n^{(-)}(a + \varepsilon)
	<  I_n^{(+)}(a ) - I_n^{(-)}(a) \;.
\end{equation}
Hence, $\partial_a I_n^{(+)}(a) -\partial_a I_n^{(-)}(a) < 0$.
Therefore, $\lambda_n \partial_a\lambda_n$ is negative.

On the other hand, if $I_n^{(+)}(a) < I_n^{(-)}(a)$ the same argument, 
after $(+) \leftrightarrow (-)$, leads to identical conclusion about the 
sign.

Since, for the Dirac field, the force in terms of the eigenvalues
$\lambda_n$ is:
\begin{equation}
	F = - \int_{-\infty}^{+\infty} \frac{d\omega}{2\pi} 
	\sum_n \frac{\lambda_n \partial_a\lambda_n}{1 - \lambda_n^2} \;,
\end{equation}
we again conclude that the force is positive, i.e., repulsive.

\section{Discussion and Conclusions}\label{sec:Conclusions}
It is appropriate to discuss here at least one example of a $\widetilde{V}$
function, in order to show how the resulting condition of the positivity
for the operator  ${\mathbb U}$ resulting from  (\ref{eq:ident}) is
realized. 
Positivity of ${\mathbb U}$ is  equivalent to positivity of both its $L$
and $R$ components. Therefore it is sufficient to consider it for a single
body.

Let us do that for a situation that has been extensively studied:
that of singular $\delta$-potentials~\cite{Fosco:2008vn}, which may be used
to enforce bag-model boundary conditions. 

A regularized version $\widetilde{V}_\varepsilon$ of $\widetilde{V}$ for a
body located at $x_1=0$, say, is: 
\begin{equation}
\widetilde{V}_\varepsilon(\omega,x_1) \,=\, \frac{g(\omega)}{\varepsilon} 
\theta(\frac{\varepsilon}{2} - |x_1|) \;,
\end{equation}
such that $\widetilde{V}_\varepsilon \to \widetilde{V}$ for $\varepsilon \to 0$.
Then, for a normalized, continuous $\psi$:
\begin{equation}
\langle \widetilde{\mathbb{U}}_\varepsilon \rangle \,=\,
\langle \psi |\widetilde{\mathbb{U}}_\varepsilon |\psi \rangle \,=\,
\big(\frac{g}{\varepsilon}\big)^2 \, + \, \frac{g}{\varepsilon} 
\big[(\psi^\dagger\gamma_1\psi)(\frac{\varepsilon}{2}) -
(\psi^\dagger\gamma_1\psi)(-\frac{\varepsilon}{2}) \big]
\end{equation}
an expression which is clearly positive for a sufficiently small
$\varepsilon$. 

For this kind of potential, one can use the natural generalization of the 
method used in~\cite{Fosco:2008vn} to prove that an odd potential leads to 
a repulsive Casimir interaction. 
As an explicit realization of this framework, we may consider the case of 
two bodies modeled by $\delta$-like potentials with opposite coupling 
constants $g$ and $-g$, separated by a distance $a$ along the $x_d$ 
direction:
\begin{equation}
	V(x)
	\;=\;
	g \left[ \delta(x_d) - \delta(x_d - a) \right] \, .
\end{equation}
This potential satisfies the required oddness condition for 
$\widetilde{V}$ and allows for an exact computation of the interaction 
energy. The result can be written as an integral over the frequency and 
momentum components perpendicular to $x_d$, denoted by $\mathbf{k}_{\perp} = 
(k_1, \ldots, k_{d-1})$. For spatial dimensions $d=1,2,3$, the vacuum 
energy is given by:
\begin{align}
	E(a)
	&\;=\;
	- \, 2^{\left\lfloor \frac{d+1}{2}\right\rfloor - 1}
	\int_{-\infty}^{\infty} \frac{d\omega}{2\pi}
	\int \frac{d^{d-1} \mathbf{k}_{\perp}}{(2\pi)^{d-1}} \,
	\log \left( 1 - \frac{16 \, g^2}{(4 + g^2)^2} \,
	e^{-2a\sqrt{\mathbf{k}_{\perp}^2 + \omega^2}} \right) \, \nonumber \\
	&\;=\;
	\frac{1}{a^d} \,
	\frac{2^{\left\lfloor 
	\frac{d+1}{2}\right\rfloor}}{(4\pi)^{\frac{d+1}{2}}} \,
	\Gamma\left(\frac{d+1}{2}\right)
	\text{PolyLog}_{d+1}\left(\frac{16 \, g^2}{(4 + g^2)^2}\right)
\end{align}
which is a positive, monotonically decreasing function of the separation 
$a$, leading to a repulsive Casimir force.

We conclude by summarizing some of the aspects of the result we present in
this article. First, we wish to remark that the property of the media being
described by an odd function is a sufficient condition for repulsion,
although not a necessary one. For instance, one also has repulsion in one of the 
systems considered in~\cite{Fosco:2024qad}, which corresponds to an atom facing a
wall which imposes bag boundary conditions. The force is repulsive, but
there is no odd function $\widetilde{V}$.

A relevant comment we wish to convey is about the relation of the 
result we present, and its dependence on the known result for bosonic 
fields: that dependence is at the level of the proof, namely, the 
formal analogy between two objects (functional determinants) is what 
makes it possible to use one result to prove the other. It is worth 
pointing out that the different statistics is what leads to repulsion 
(rather than attraction) when relating both functional determinants.

We also wish to elaborate on another important aspect of the approach 
we followed: note that we have expressed the Casimir energy for a 
Dirac field in the presence of space-dependent mass (the function 
$\widetilde{V}$ in (\ref{eq:dirop}) ) from the determinant of a 
fluctuation operator ${\mathcal H}$, defined as follows:
\begin{equation}
 {\mathcal H} \equiv {\mathcal D}^\dagger {\mathcal D} \;\;,\;\;
 {\mathcal D} \;=\; \not \! \nabla + i \gamma_0  \omega 	+ 
 \widetilde{V} \;.
\end{equation}
However, one could certanly have considered the operator 
${\mathcal H}' \equiv {\mathcal D} {\mathcal D}^\dagger$, which has the 
same spectrum as ${\mathcal H}$, except for the 
existence of  zero modes. Indeed, that is the content of the 
Atiyah–Patodi–Singer index theorem~\cite{AtiyahI,AtiyahII,AtiyahIII} 
which can be used to relate the spectral asymmetry between ${\mathcal 
H}$ and ${\mathcal H}'$ to topological properties of the Dirac 
operator.  
In cases where such asymmetry exists, in order to evaluate the 
contribution of the fermionic modes, one should take the average of the 
contributions to the energy due to the operators ${\mathcal H}$ and 
${\mathcal H}'$~\cite{Chan,Graham,Dunne}. For the 
system we are considering, that would correspond to taking the average 
 of the energies due to two systems, defined with two values of 
the object defined in (\ref{eq:ident}), namely: 
$\widetilde{\mathbb{U}}^{\pm} = \widetilde{V}^2 \pm (\not \! \nabla 
\widetilde{V})$. Equivalently, one should take the average of the 
energies resulting from $L$ and $R$ in there original positions, and 
the energy when they are reflected.
   For the kind of background we are considering, a 
space-dependent mass could produce zero modes for domain-wall mass 
profiles and in an odd number of spacetime dimensions, as in the Callan 
and Harvey mechanism~\cite{Callan:1984sa}. 
Note, however, that that would require a rather different kind of 
configuration: the mass we are considering here 
has two spatial lumps (since we are dealing with a Casimir interaction 
phenomenon) and it vanishes except for the region occupied by the 
bodies.

We conclude by pointing out an interesting problem regarding this kind
of result, about the sign of the Casimir interaction:
for the bosonic version, it would be to know how
far a departure from a perfectly symmetric configuration can go without
changing the conclusion, i.e., attractive force. And its equivalent for the
Dirac field case and departures from a perfectly odd configuration.

\section*{Acknowledgments}
The authors thank ANPCyT, CONICET and UNCuyo for financial support.

\end{document}